# Small witnesses, accepting lassos and winning strategies in $\omega$-automata and games[*]

Rüdiger Ehlers

Reactive Systems Group, Department of Computer Science, Saarland University
66123 Saarbrücken, Germany

Obtaining accepting lassos, witnesses and winning strategies in $\omega$-automata and games with $\omega$-regular winning conditions is an integral part of many formal methods commonly found in practice today. Despite the fact that in most applications, the lassos, witnesses and strategies found should be as small as possible, little is known about the hardness of obtaining small such certificates. In this paper, we survey the known hardness results and complete the complexity landscape for the cases not considered in the literature so far. We pay particular attention to the approximation hardness of the problems as approximate small solutions usually suffice in practice.

## 1 Introduction

Automata and games with $\omega$-regular acceptance or winning conditions have been proven to be valuable tools for the construction and analysis of complex systems and are suitable computation models for logics such as the monadic second-order logic of one or two successors [11, 23, 1, 24]. By reducing a decision problem to determining the winning player in a game or checking emptiness of an automaton, the algorithmic aspect of solving the problem can easily be separated from the details of the application under concern. Winning strategies for one of the players in a game or accepting runs in an automaton in turn can be used as *certificates* for the answer to the original problem.

In many applications, the actual content of such a certificate is also of interest. In the game context, for example, when synthesizing finite state systems [25, 18] from temporal logic specifications, the certificate for the system player in the corresponding game represents a system that satisfies the specification. In alternating time logic (ATL) [1], the question is imposed whether agents in a certain setting can ensure certain global properties of a system. A certificate for the corresponding model checking game represents either a set of strategies for the agents to achieve their goal or a counter-strategy for the remaining agents to prevent this. In the automaton context, on the other hand, when model checking a finite-state system against some linear-time specification, the question whether there exists an erroneous computation of the system is typically reduced to the emptiness problem of some automaton and an accepting run of the automaton then is a finite representation of such a computation and a proof that it violates the specification at the same time.

In all these cases, certificates (the winning strategies in the game or the accepting runs of the automaton) that have a smaller representation are normally preferred. Such solutions are easier to comprehend and have (computational) advantages if used in successive steps (like building circuits from strategies in a synthesis game) or for analysing why a certain property holds or not. While in the automaton context, there exist some results for Büchi and generalised Büchi acceptance conditions, little research has been performed on obtaining small strategies in games, even though it has been observed that needlessly large strategies are often problematic [17, 2].

---

[*]This work was supported by the German Research Foundation (DFG) within the program "Performance Guarantees for Computer Systems" and the Transregional Collaborative Research Center "Automatic Verification and Analysis of Complex Systems" (SFB/TR 14 AVACS).

From a technical point of view, two types of game strategies have to be considered: *arena-based* and *stand-alone* ones. In the former case, the behaviour of a player at any point in time is dependent on the game position she is in at that point. A typical representative of this class are the *positional strategies*, which only depend on the game position. These for example give insight on why a modal $\mu$-calculus formula is valid in some model or provide information about why a specification is unrealisable in synthesis, as the obligations are encoded into the game graph and are thus easily interpretable. In case of the more complex generalised Büchi, Streett and Muller winning condition types, in which *memoryless determinancy* (which guarantees that whenever some player has a winning strategy, she also has a positional strategy) does not hold, so-called *finite memory* strategies are considered instead, in which the player always decides the next memory content and the next move depending on the current memory content and the current position in the game. Here, again, the current position in the game arena is taken into account for a move.

On the other hand, stand-alone strategies do not use the current position in the game as an external input, and must use their memory to keep track of information about the history of the play, if needed. They are typically represented as Mealy or Moore automata that read the actions of one player and output the actions of the other player such that the other player wins the game when simulating the behaviour of the automaton. These are, for example, a suitable model for synthesis using realisable specifications as such strategies are typically smaller than positional ones but sufficient for transforming the strategy into a circuit that satisfies the specification. While they are equi-expressive as finite-memory strategies, they give rise to a different definition of strategy size, which motivates their consideration here.

As automata can be seen as games with only one player, the definition of these strategy types directly carries over to the automaton case. Here, however, arena-based strategies are typically called *accepting lassos* in the literature. Likewise, winning stand-alone strategies for one-player games have been introduced under the name *witnesses* [17], which are accepted words of the form $uv^\omega$ for some finite words $u$ and $v$. In the model checking context, an accepting lasso typically represents an erroneous computation of a system and a proof for the violation of the property at the same time. Witnesses are used to represent only the computation in a concise way, which is often more useful as the accepting lassos incorporate information about the specification automaton cycle corresponding to the violation, which the user of a model checking system typically does not care about.

For a positional strategy, we denote by its size the number of positions in the game that can be reached on some path when the corresponding player follows the strategy from the initial position. For a lasso, its size is defined as the length of its cycle plus the length of the path to the cycle. For finite memory strategies, their size is defined to be the number of combinations of memory content and game positions that can occur along some path when following the strategy from the initial game position (and the initial memory content). For stand-alone strategies in two-player games, the number of states of the corresponding Moore automaton is taken as the size measure. The size of a witness $uv^\omega$ is defined to be $|u| + |v|$.

Using these definitions, the sizes of the smallest finite-memory strategy and the smallest positional strategy coincide in game types that are memoryless determined (which is proven later in Theorem 1).[1] Thus, we henceforth simplify our presentation by using the term arena-based strategy for both positional and finite-memory strategies as no cases remain that could cause ambiguity. The definitions above make sure that the size measures for lassos and witnesses are special cases of the size measures for arena-based and stand-alone strategies, respectively, which stresses the conceptual similarities of the one- and two-player cases.

For automata, the problem of finding small non-emptiness certificates has been tackled by a couple of researchers in the context of model checking. In case of the Büchi acceptance condition, it is well-known that shortest accepting lassos can be computed in polynomial time (see, e.g., [10, 22]). On the other hand, for automata with the generalised Büchi acceptance condition, the problem is known to be NP-complete [5]. For finding shortest witnesses, the problem is known to be NP-complete even for the Büchi acceptance condition [17].

Despite these results, the complexity landscape for the problems of finding shortest lassos and witnesses in $\omega$-automata for the various acceptance condition types is not completely known yet. In the case of two-player games, the situation is a little bit better: here it is known that finding a minimal positional strategy in a safety game is an NP-complete problem and there also exists no polynomial-time approximation scheme (PTAS) [4]. As safety games are a special case of parity, Büchi, co-Büchi, generalised Büchi, Rabin and Streett games, the NP-hardness result also applies to these cases.

---

[1] As it is questionable to examine the problem of determining the size of the smallest positional strategy for games with winning condition types that do not induce memoryless determinancy, we may restrict our attention to finite-memory and stand-alone strategies for these.

As for practical applications, it normally suffices to give small but not necessarily minimal winning strategies (such as, e.g., strategies of size not more than two times the size of the smallest strategy), accepting lassos or witnesses, such approximation hardness results are as important as solutions to the exact hardness problems. Nevertheless, again, such questions have only been answered for some special cases so far.

## 1.1 Scope of this paper

In this paper, we perform a rigorous study of the problems of finding small accepting lassos, witnesses and winning strategies in $\omega$-automata and games for all commonly used acceptance/winning condition types. For all of these cases, we review the known complexity results for the exact minimization of the size of accepting lassos, witnesses or winning strategies and complete the complexity landscape for the cases not considered in the literature so far. This results in the first complete exposition of the borderline between the hard and simple problems in this context. Directly connected with this survey is the question how well the size of the minimal winning strategies, witnesses and accepting lassos can be approximated in polynomial time (provided that P$\neq$NP), as only then, the problem becomes hard for practice as well. The results we obtain in this paper for the approximability of the problems considered are mostly negative: for example, we show that approximating the minimal witness size within any polynomial is NP-complete even for an automaton having the simple safety acceptance condition. On the positive side, we also provide a general exponential-quality approximation scheme that is applicable to many of the problem settings discussed in this paper.

Table 1 on page 8 contains a summary of the results. This technical report complements an earlier paper by its author [9] that only considered the one-player case and appeared at the 4th International Conference on Language and Automata Theory and Applications (LATA), 2010. In this technical report, we complete the picture for the two-player case and reuse hardness results from [9] in those sub-cases in which they are applicable.

This paper is structured as follows: In the next section, we define game arenas, acceptance and winning conditions and the two notions of strategies used in this work. A discussion of the corresponding definitions for the one-player case and a summary of the terminology for approximation algorithms closes the preliminaries section. Then, we give an overview of the results. We proceed by showing that all problems discussed in this paper are contained in the complexity class NP, followed by a novel approximation hardness result for obtaining small arena-based strategies. A conclusion sums up the results of this paper and provides an outlook.

## 1.2 Related work

To the best of our knowledge, the amount of related work for finding small winning strategies, witnesses and accepting lassos in one- and two-player games in addition to the works aforementioned is very limited. Holtmann and Löding [13] describe a scheme that typically allows for finding smaller strategies in a weak Muller game by converting the game to a weak parity game and using an automaton minimisation technique such that winning strategies in this setting are usually smaller. Furthermore, Bloem at al. [2] describe a symbolic strategy selection heuristic in a game that has already been pruned from all positions that are losing for the first player. The heuristic is applied to synthesizing circuits from a synthesis game. They report that the sizes of the synthesized circuits are usually huge compared to hand-made implementation satisfying the same specifications. In [16], Kupferman et at al. show that checking if there exists a stand-alone strategy of a given size in a Büchi game is NP-complete.

## 2 Preliminaries

We start by recalling the basic definitions used for two-player games and point out the differences to one-player games (automata). Then, we formally define the strategy size concepts and recall the basic terminology for approximation algorithms.

## 2.1 Games

A game arena $\mathcal{G} = (V_0, V_1, A_0, A_1, E_0, E_1, v_0)$ is defined with two finite sets of vertices (also called positions) $V_0$ and $V_1$ for the two players 0 and 1 of the game, the finite sets of actions they can choose from $(A_0, A_1)$, their possible moves $(E_0 : V_0 \times A_0 \rightharpoonup V_1, E_1 : V_1 \times A_1 \rightharpoonup V_0)$ and the initial position $v_0 \in V_0$. For all $p \in \{0, 1\}$, we say that some action $a \in A_p$ is available to player $p$ at a position $v \in V_p$ if $E_p(v, a)$ is defined. We assume that all of these sets are finite. A game position $v \in V_p$ for some $p \in \{0, 1\}$ is called a successor position of some position $v' \in V_{1-p}$ if there exists some action $a \in A_{1-p}$ such that $E_{1-p}(v', a) = v$. A game is also called a one-player game for some player $p \in \{0, 1\}$ if we have $|A_{1-p}| = 1$ and $E_{1-p}$ is a total function. One-player games for player 0 are also called $\omega$-word automata or simply $\omega$-automata if for all $v \in V_1$, $E_1(v_1, a)$ is defined for $a \in A_1$. In this case, the identifiers $V_1$, $A_1$ and $E_1$ are typically omitted from the game definition and for all $a \in A_0$ and $v \in V_0$, we take $E_0'(v, a) = E_1(E_0(v, a), a')$ for the only element in $a' \in A_1$ as the modified edge function for player 0 (which circumvents visiting vertices in $V_1$ altogether). Here, the vertices are also called states, the action set is called the alphabet and $E_0$ is called the transition relation. The state set in this case is typically referenced to by the letter $Q$ whereas $\Sigma$, $\delta$ and $q_0$ represent the alphabet, the transition relation and the initial state, respectively.

We say that there exists some path from some vertex $v \in V_0$ to some vertex $v' \in V_0$ if for some $a_1^0, \ldots, a_n^0 \in A_0$ and $a_1^1, \ldots, a_n^1 \in A_1$, we have $E_1(E_0(\ldots E_1(E_0(v, a_1^0), a_1^1), \ldots), a_n^1) = v'$. We define the size of a game arena as $|\mathcal{G}| = |V_0| + |V_1| + |A_0| + |A_1| + |\{(v, a) \in V_0 \times A_0 \mid E_0(v, a) \text{ is defined}\}| + |\{(v, a) \in V_1 \times A_1 \mid E_1(v, a) \text{ is defined}\}|$.

A decision sequence in $\mathcal{G}$ is a sequence $\rho = \rho_0^0 \rho_0^1 \rho_1^0 \rho_1^1 \ldots$ such that for all $i \in \mathbb{N}_0$, $\rho_i^0 \in A_0$ and $\rho_i^1 \in A_1$. A decision sequence $\rho$ induces an infinite play $\pi = \pi_0^0 \pi_0^1 \pi_1^0 \pi_1^1 \ldots$ with $\pi_0^0 = v_0$ and for all $i \in \mathbb{N}_0$ and $p \in \{0, 1\}$, we have $E_p(\pi_i^p, \rho_i^p) = \pi_{i+p}^{1-p}$. Likewise, a finite play $\pi = \pi_0^0 \pi_0^1 \pi_1^0 \pi_1^1 \ldots \pi_n^p$ is induced for some $p \in \{0, 1\}$ if $\pi_0^0 = v_0$, for all $i \in \mathbb{N}_0$ and $p' \in \{0, 1\}$, $i < n + p \cdot (1 - p')$ implies $E_{p'}(\pi_i^{p'}, \rho_i^{p'}) = \pi_{i+p'}^{1-p'}$, and $E_p(\pi_n^p, \rho_n^p)$ is undefined. For such a play, we say that $\pi$ ends in $\pi_n^p$.

In a game, the players involved try to win the game according to the winning condition imposed. A play ending in a vertex in $V_0/V_1$ is winning for player 1/0, respectively. Within the scope of this paper, we always view the game from player 0's view, so we simply call a play winning or accepting if it is winning for player 0. For an infinite play $\pi$, whether $\pi$ is winning is defined over the vertices of $V_0$ occurring infinitely often along the play (denoted by $\inf_{V_0}(\pi)$):

- For a *safety winning condition*, player 0 wins on every infinite play.

- For a *Büchi winning condition* $\mathcal{F} \subseteq V_0$, $\pi$ is accepting if $\inf_{V_0}(\pi) \cap \mathcal{F} \neq \emptyset$.

- For a *co-Büchi winning condition* $\mathcal{F} \subseteq V_0$, $\pi$ is winning if $\inf_{V_0}(\pi) \cap \mathcal{F} = \emptyset$.

- For a *generalised Büchi winning condition* $\mathcal{F} \subseteq 2^{V_0}$, $\pi$ is accepting if for all $F \in \mathcal{F}$, $\inf_{V_0}(\pi) \cap F \neq \emptyset$.

- For a *parity winning condition*, $\mathcal{F} : V_0 \to \mathbb{N}$ and $\pi$ is accepting if $\max\{c(v) \mid v \in \inf_{V_0}(\pi)\}$ is even.

- For a *Rabin winning condition* $\mathcal{F} \subseteq 2^{V_0} \times 2^{V_0}$, $\pi$ is accepting if for $\mathcal{F} = \{(F_1, G_1), \ldots, (F_n, G_n)\}$, there exists some $1 \leq i \leq n$ such that $\inf_{V_0}(\pi) \subseteq F_i$ and $\inf_{V_0}(\pi) \cap G_i \neq \emptyset$.

- For a *Streett winning condition* $\mathcal{F} \subseteq 2^{V_0} \times 2^{V_0}$, $\pi$ is accepting if for $\mathcal{F} = \{(F_1, G_1), \ldots, (F_n, G_n)\}$ and for all $1 \leq i \leq n$, we have $\inf_{V_0}(\pi) \not\subseteq F_i$ or $\inf_{V_0}(\pi) \cap G_i = \emptyset$.

- For a *Muller winning condition* $\mathcal{F} \subseteq 2^{V_0}$, $\pi$ is accepting if $\inf(\pi) \in \mathcal{F}$.

We define that player 1 wins all infinite plays that are not winning for player 0. Given some game arena $\mathcal{G}$ and some *acceptance component* $\mathcal{F}$ of type $t \in \{\text{Safety, Büchi, co-Büchi, generalised Büchi, parity, Rabin, Streett, Muller}\}$, we say that $\langle \mathcal{G}, \mathcal{F} \rangle$ is a *game* of type $t$ (for safety games, the $\mathcal{F}$ symbol is usually omitted). We say that a play is winning in $\langle \mathcal{G}, \mathcal{F} \rangle$ if and only if it is winning for $\mathcal{F}$ using the definition of $t$-type acceptance given above. By abuse of notation, we henceforth add the acceptance component to the game arena tupel. We define the size of a game $\langle \mathcal{G}, \mathcal{F} \rangle$ to be $\langle \mathcal{G}, \mathcal{F} \rangle = |\mathcal{G}| + |\mathcal{F}|$.[2]

---

[2] As in this paper, we are only interested in the borderline between NP-complete problems and those that are in P (assuming P$\neq$NP), we can safely ignore the fact that an explicit encoding of $\mathcal{F}$ might actually be slightly bigger.

Given some $t$-type game $(V_0, V_1, A_0, A_1, E_0, E_1, v_0, \mathcal{F})$, a strategy for player 0 is a function $f : (A_0; A_1)^* \to A_0$, and a strategy for player 1 is a function $f : (A_0; A_1)^* A_0 \to A_1$, both mapping prefix decision sequences to an action to be chosen next. A decision sequence $\rho = \rho_0^0 \rho_0^1 \rho_1^0 \rho_1^1 \ldots$ is said to be in correspondence to a strategy $f$ for player $p$ if for every $i \in \mathbb{N}_0$, we have $\rho_n^p = f(\rho_0^0 \rho_0^1 \ldots \rho_{n+p-1}^{1-p})$. A strategy is winning for player $p$ if all plays in the game that are induced by some decision sequence that is in correspondence to $f$ are winning for player $p$. It is a well-known fact that for the winning condition types given above, for all possible game arenas, there exists a winning strategy for precisely one of the players [11]. Winning strategies are also called *accepting* in the word automaton case. We call a state $v \in V_0$ winning for player $p$ if changing the initial state to $v$ makes or leaves the game winning for player $p$. Likewise, a state $v' \in V_1$ is called winning for player $p$ if a modified version of the game that results from introducing a new initial state with only one transition to $v'$ is (still) winning for player $p$.

## 2.2 Strategy types

In this this work, we are concerned with positional, finite-memory and stand-alone strategies. These all have finite representations which induce strategy functions $f$ of the general type stated above. By abuse of terminology (and following the commonly used definitions in the literature) we call both the finite representation as well as $f$ a strategy.

**Positional strategies:** A positional strategy for player $p$ in a game $(V_0, V_1, A_0, A_1, E_0, E_1, v_0, \mathcal{F})$ is finitely represented by a partial function $f' : V_p \rightharpoonup A_p$ that maps each vertex of the player to an action that is chosen by the player whenever the vertex is visited. The general strategy $f$ induced by $f'$ is defined as:

$$f(\rho_0^0 \rho_0^1 \ldots \rho_n^{1-p}) = f'(E_{1-p}(\ldots E_1(E_0(v_0, \rho_0^0), \rho_0^1), \ldots, \rho_n^{1-p}))$$

In order to assure that $f$ is well-defined, we require that for every position $v \in V_p$ that is in the domain of $f'$ and every action $a \in A_{1-p}$ of the other player, $E_{1-p}(E_p(v, f(v)), a)$ is also in the domain of $f$. If a game arena is of safety-, Büchi-, co-Büchi- or parity-type, then there exists a winning positional strategy for a player if and only if there exists some winning strategy for the same player [11]. This fact also holds for player 0 in Rabin games and player 1 in Streett games. Given a positional strategy $f' : V_p \rightharpoonup A_p$, by $|f'|$ we denote the size of the domain of $f'$ (i.e., the size of the subset of $V_p$ for whose elements $e$, $f'(e)$ is defined).

**Finite-memory strategies:** A finite-memory strategy for player $p$ in a game $(V_0, V_1, A_0, A_1, E_0, E_1, v_0, \mathcal{F})$ is finitely represented by a tuple $\mathcal{S} = (S, s_0, f', f'_M)$ for some state set $S$, some functions $f' : S \times V_p \rightharpoonup A_p$ and $f'_M : S \times V_p \to S$ and some initial state component $s_0$. For $p = 0$, we have $s_0 \in S$ whereas for $p = 1$, we have $s_0 : V_1 \to S$ (such that the initial move of player 0 can be taken into account for the initial memory content).

Intuitively, a finite-memory strategy has some finite memory with domain $S$ that it updates using the function $f'_M$ and the moves of the other player. The behaviour of the player $p$ then not only depends on the current position as in positional strategies, but also on the memory content. The general strategy $f$ induced by $\mathcal{S}$ is defined as:

$$f(\rho_0^0 \rho_0^1 \ldots \rho_{n+p-1}^{1-p}) = f'(f''(\rho_0^0 \rho_0^1 \ldots \rho_{n+p-2}^{1-p}), E_{1-p}(\ldots E_1(E_0(v_0, \rho_0^0), \rho_0^1), \ldots, \rho_{n+p-1}^{1-p}))$$

for the auxiliary function

$$f''(\rho_0^0 \rho_0^1 \ldots \rho_{n+p-1}^{1-p}) = f'_M(f''(\rho_0^0 \rho_0^1 \ldots \rho_{n+p-2}^{1-p}), E_{1-p}(\ldots E_1(E_0(v_0, \rho_0^0), \rho_0^1), \ldots, \rho_{n+p-1}^{1-p}))$$

with $f''(\epsilon) = s_0$ if $p = 0$ and $f''(\rho_0^0) = s_0(E_0(v_0, \rho_0^0))$ if $p = 1$.

Given such a finite-memory strategy, we define its size to be the size of the domain of $f'$.

**Stand-alone strategies:** Finally, a stand-alone strategy for player 0 is defined as a Moore automaton [20] $\mathcal{M} = (Q, A_1, A_0, \delta, L, q_0)$ with some finite set of states $Q$, some initial state $q_0 \in Q$, an input symbol set $A_1$, an output symbol set $A_0$, some transition function $\delta : Q \times A_1 \to Q$ and a labelling function $L : Q \to A_0$.

The corresponding general strategy is defined by $f(\rho_0^0 \rho_0^1 \ldots \rho_n^1) = L(\delta(\ldots \delta(q_0, \rho_0^1), \ldots, \rho_n^1))$. The definition for player 1 is equivalent, except that the automaton model is then of Mealy [20] type. The size of a finite memory strategy is defined to be the number of states in the Moore/Mealy automaton.

## 2.3 A note on the definitions of games and strategies

Note that the definitions of the concepts given above differ between the works in the literature. However, except for the Muller winning condition, the game model given above allows a polynomial re-encoding into the other commonly used models (where, for example, the positions of player 1 that occur infinitely often along runs are also taken into account for determining whether a play is winning, or there is a set of initial positions instead of a single position given, etc.), making the hardness proofs and approximation algorithms in the next sections valid for these models as well.

## 2.4 Strategy finding algorithms and their approximation quality

Given a game graph $\mathcal{G}$ and some value $n \in \mathbb{N}$, the decision version of finding small winning arena-based or stand-alone strategies, accepting lassos or witnesses (which are accepting by definition) is to decide whether there exists a winning arena-based or stand-alone strategy, accepting lasso or witness of size at most $n$ or not. We say that some algorithm approximates one of these problems within some function $g : \mathbb{N} \to \mathbb{N}$ if it always returns **false** whenever there does not exist a strategy of size $g(n)$ and returns **true** if there exists a strategy of size at most $n$. In the other cases, the result of the algorithm is unconstrained.

The construction version of the problem is defined likewise. Given a game graph $\mathcal{G}$, the construction version of finding small arena-based or stand-alone strategies, lassos or witnesses is to decide whether there exists an arena-based or stand-alone strategy, lasso or witness of size (at most) $n$ and to construct such a strategy/lasso/witness if it exists. We say that some algorithm approximates one of these problems within some function $g : \mathbb{N} \to \mathbb{N}$ if it always returns that no such strategy exists whenever there does not exist a strategy/lasso/witness of size $g(n)$ and returns a strategy of size at most $g(n)$ if there exists one of size $n$. In the remaining cases, the algorithm may either return a valid winning strategy/lasso/witness of size not more than $g(n)$ or return no such certificate.

We say that an algorithm approximates the minimal strategy/lasso/witness size within some constant $c$ if it approximates the minimal strategy/lasso/witness size within $g(n) = c \cdot n$. For the hardness and non-approximability results, we assume that P$\neq$NP (otherwise the decision versions of all problems discussed in this paper are solvable in polynomial time).

## 2.5 On positional and finite-memory strategies

In the introduction, we justified grouping positional and finite-memory strategies into arena-based strategies by the claim that for game types that are memoryless determined, adding memory to strategies does not change the size of the smallest winning strategy in a game. We say that a game type is memoryless determined if for every game with such a winning condition, the existence of a winning strategy implies the existence of a winning positional strategy.

**Theorem 1.** *In Rabin games and their special cases, the sizes of the smallest positional strategy for player* 0 *and the smallest finite-memory strategy for player* 0 *coincide.*

*Proof.* Let $\mathcal{A} = (V_0, V_1, A_0, A_1, E_0, E_1, v_0, \mathcal{F})$ be a Rabin game and $\mathcal{S} = (S, s_0, f', f'_M)$ be a finite-memory strategy that is winning for player 0. As $\mathcal{S}$ is winning, it makes sure that along a play from $v_0$ that is in correspondence to $\mathcal{S}$, only vertices that are winning for player 0 can be visited. As a consequence, we can assume without loss of generality that the domain of $f'$ contains only vertices of $\mathcal{A}$ that are winning for player 0 (as otherwise we can simply drop non-winning positions from the domain).

Now consider a game $\mathcal{A}'$ that corresponds to the restriction of $\mathcal{A}$ to the vertices that can be visited along some path from the initial vertex of $\mathcal{A}$ when player 0 follows the strategy $\mathcal{S}$. Since $\mathcal{S}$ is winning, $\mathcal{A}'$ is winning and the number of player 0-vertices in $\mathcal{A}'$ is $c = |\{q \in V_0 | \exists s \in S : f'(s, q) \text{ is defined}\}|$. However, as $\mathcal{A}'$ is still a Rabin game (or of a type that is a special case of Rabin acceptance), which always permit memoryless winning strategies, it is also guaranteed that $\mathcal{A}'$ has a memoryless strategy of size at most $c \leq |f'|$. As such a strategy can be projected back to a corresponding strategy in $\mathcal{A}$, the claim follows. □

As a consequence, when examining the hardness of the problems of finding smallest arena-based strategies in games, it suffices to only consider positional strategies for Rabin games and their special cases, and for the remaining types, for which memoryless determininancy does not hold, to use finite-memory strategies.

It should be noted that from a theoretical point of view, for the approximation hardness of the problems, this conclusion cannot be drawn as allowing memory might make approximating the size of the smallest winning strategy easier. However, in the hardness proofs in this paper and in [9], the reductions performed immediately give rise to the insight that this is not the case.

## 3 An overview of the results

The problems of finding short lassos in automata and small strategies in games discussed in this paper are all contained in the complexity class NP (which is proven in Section 4). For Streett games, which are co-NP-complete to solve, and Muller games, which require an exponential amount of memory in general [8], we assume a unary encoding of the strategy size to make the problem be contained in NP.

It remains to answer the question which (approximation) problems in this context are NP-complete and which are solvable in polynomial time. Many of the theorems cited or given in the following sections to answer this question have implications for multiple game/automaton models discussed here. For example, hardness results for the one-player case carry over to the two-player case. Also, hardness results for one winning condition type carry over to the more general winning condition types that do not require a super-polynomial re-encoding. Likewise, algorithms for game solving can also be used for special cases. Thus, to aid the reader in seeing all consequences of the individual theorems, we give an overview of the results here and provide Table 1 that surveys all results.

### 3.1 Summary of the results in [9]

We start by re-stating the results in [9] that contribute to the overview below.

**Theorem 2** ([9], Section 3.1). *Finding shortest accepting lassos in Rabin automata is doable in polynomial time.*

**Theorem 3** ([9], Theorem 2). *The size of a shortest accepting lasso in generalised Büchi automata is NP-hard to approximate within any constant.*

**Theorem 4** ([9], Theorem 4). *The size of a shortest accepting lasso in Muller automata is NP-hard to approximate within $\frac{321}{320} - \epsilon$ for every $\epsilon > 0$.*

**Theorem 5** ([9], Theorem 5). *Finding an accepting lasso of size not more than $\lceil \log_2(|F_1|) \rceil$ times the length of a shortest accepting lasso for the largest element $F_1$ of the acceptance component in a Muller automaton is possible in polynomial time.*

**Theorem 6** ([9], Theorem 6 and its discussion). *The size of a shortest witness in a safety or Muller automaton is NP-hard to approximate within any polynomial function.*[3]

**Theorem 7** ([9], Theorem 7). *Let $c > 1$ and $\Sigma$ be some fixed finite alphabet. Given some $\omega$-automaton $\mathcal{A} = (Q, \Sigma, q_0, \delta, \mathcal{F})$ with any of the acceptance types considered in this paper, computing a word $uv^\omega$ such that $|u|+|v|$ is not longer than $c^n$ for $n$ being the minimal witness length can be done in time polynomial in $|\mathcal{A}|$.*

### 3.2 New results in this report

We add the following hardness results to the aforementioned facts of [9] in this report:

---

[3]Note that in principle, Theorem 6 from [9] does not directly apply as in [9], the automata are non-deterministic whereas the 1-player games in this report correspond to deterministic automata. However, the automaton built in Theorem 6 of [9] is actually deterministic and thus the Theorem is applicable here as well. The other results are not affected by this mismatch as they are concerned with shortest lassos (for which non-determinism does not matter) or are approximability results, for which determinism can only make the situation easier.

Table 1: Overview of the complexity results of the problems of finding small winning strategies, accepting lassos and witnesses for games and automata. All problems under consideration are contained in NP (for Streett and Muller two-player games, we assume a unary encoding of the strategy size). In this table, we only consider polynomial-time algorithms and assume that P≠NP. The symbol $Q$ denotes the state set of an automaton under concern.

| | | One-player games | Two-player games |
|---|---|---|---|
| **Arena-based strategies** | **Safety, Büchi, co-Büchi and parity winning cond.** | solvable precisely in polynomial time | not approximable within any constant, approximable within every exponential function for fixed action sets |
| | **Rabin winning cond.** | | not approximable within any strictly increasing function |
| | **Generalised Büchi winning cond.** | not approximable within any constant, approximable within every exponential function for a fixed alphabet | not approximable within any constant |
| | **Streett winning cond.** | | |
| | **Muller winning cond.** | not approximable within $\frac{321}{320} - \epsilon$, approximable within $\lceil \log_2 |Q| \rceil$ | |
| **Stand-alone strategies** | **Safety, Büchi and co-Büchi winning cond.** | not approximable within any polynomial, approximable within every exponential function for a fixed alphabet | not approximable within any polynomial, approximable within every exponential function for fixed action sets |
| | **Generalised Büchi, Parity, Rabin, Streett and Muller winning cond.** | | not approximable within any polynomial |

- **Theorem 11:** The size of a smallest winning arena-based strategy in a safety or Muller game is NP-hard to approximate within any constant.

- **Corollary 12:** For any strictly increasing function $g : \mathbb{N}_0 \to \mathbb{N}_0$, approximating the size of a smallest winning arena-based strategy in Rabin games within $g$ is NP-hard.

We also contribute an approximability result:

- **Corollary 14:** Approximating the size of a smallest winning arena-based strategy for player $0$ in parity games with a constant number of colours (and their special cases) within any exponential function can be performed in time polynomial in the size of the game if the sets of actions for both players are fixed.

## 4 Finding small winning strategies in games is in NP

In this section, show that the problem of finding small winning strategies for various winning condition types is contained in NP. We consider the decision versions of the problems here, i.e., given a bound $b$ and a game $\mathcal{G}$, we check if there exists a strategy for player $0$ in $\mathcal{G}$ that is winning and has a size of at most $b$.

In fact, for showing that this problem is in NP (i.e., that it can be solved by a non-deterministic Turing machine in polynomial time), it suffices to show that checking whether a given strategy is valid is solvable in polynomial time (in the size of the game) as, given a bound $b$, the strategy can be guessed by the non-deterministic Turing machine in time polynomial in $b$. For Streett and Muller winning conditions in two-player games, we assume a unary encoding of the bound $b$ in order to make sure that the number of bits that have to be guessed by the non-deterministic Turing machine is only polynomial in the input size. As for the other game types (safety, Büchi, co-Büchi, parity, Rabin and generalised Büchi two-player games, and Streett and Muller one-player games), there exist trivial upper bounds on the sizes of minimal strategies (for generalised Büchi winning conditions in two-player games: the number of positions in the game times the number of acceptance sets; for the remaining acceptance types in two-player games, the number of positions in the game; for Streett and Muller automata: at most the square of the number of states) that are only polynomial in the size of the input, we do not need to restrict the encoding of $b$ in order to obtain a non-deterministic algorithm that runs in time polynomial in the input size. Note that due to co-NP-completeness of Streett game solving and the fact that winning strategies in a Streett game might require an exponential amount of memory [8], a non-binary encoding is necessary for the correctness of the claim unless co-NP=NP.

For checking whether the guessed strategies are correct, we use the following simple facts:

- Given a two-player game and a positional strategy $f'$ for player $0$, it can easily be checked whether the strategy is structurally valid, i.e., if for every $v \in V_0$ reachable on some play in correspondence with $f'$, $f'(v)$ and $E_0(v, f'(v))$ are defined. Also, the computation of the product of the game and the strategy that results from stripping all choices for player $0$ that do not correspond to the strategy is easy. Note that the resulting game is a one-player game for player $1$.

- Given a two-player game and a finite-memory strategy for player $0$, we can compute a one-player game in which player $0$'s decisions are fixed by taking the product state space of the game graph and the memory automaton in time polynomial in the size of the game and the memory automaton. This *product automaton* contains a non-accepting play if and only if player $0$'s strategy is not winning.

- Given a two-player game and a stand-alone strategy for player $0$, we can compute a one-player game in which player $0$'s decisions are fixed by taking the product state space of the game graph and player $0$'s strategy automaton in time polynomial in the size of the game and the strategy. This product automaton contains a non-accepting play if and only if player $0$'s strategy is not winning.

- For every reachable vertex in a game, we can check for the existence of a cycle that contains the position and is winning for player $1$ in polynomial time. By duality, for Rabin winning conditions, this follows from the existence of efficient emptiness checking algorithms for Streett word automata (see, e.g., [12]). Likewise, for Streett winning conditions, we check for Rabin automaton emptiness, which can also be done in polynomial time (see Theorem 2). For the simpler winning conditions, the respective result follows from the results for Rabin and Streett winning conditions (with only polynomial blow-up for changing the representations accordingly). For Muller winning conditions, we can use the recently found polynomial algorithm for solving Muller games [14] to check if the game is winning for player $1$.

So, a (non-deterministic) algorithm for solving the problem of checking if there exists some arena-based strategy of size $b$ can be given as follows:

1. Guess some finite-memory strategy that has the memory state set $\{1, \ldots, b\}$.

2. Check that the strategy has size $b$. If this is not the case, return "no" as the answer. Otherwise build the product game.

3. For Muller games, use the polynomial algorithm for solving the game to check if the guessed strategy is winning. Otherwise, interpret the resulting game as word automaton (as the choices of one player have already been resolved) and check for emptiness for the dual acceptance condition. If and only if its language is empty, return "yes" as the answer.

For stand-alone strategies, the problem can be solved as follows:

1. Guess some suitable Moore automaton $\mathcal{A}$ that represents the strategy with $b$ states.

2. Compute the product game of $\mathcal{A}$ and the original game graph that is empty if and only if $\mathcal{A}$ is winning for player 1.

3. Like in the arena-based strategy case, return "yes" if and only if the resulting one-player game is winning for player 0.

It remains to be shown how the product games can be computed. We only give the construction for the case of stand-alone strategies and Rabin games here. The other cases can be performed analogously.

**Lemma 8.** *Given some Rabin-type game $\mathcal{G} = (V_0, V_1, A_0, A_1, E_0, E_1, v_0, \mathcal{F})$ and a stand-alone strategy $M = (Q, A_1, A_0, \delta, L, q_0)$, we can construct a one-player Rabin game $\mathcal{G}' = (V_0', V_1', A_0, A_1, E_0', E_1', v_0', \mathcal{F}')$ that is winning for player 0 if and only if $M$ is a winning strategy in $\mathcal{G}$ as follows:*

- $V_0' = Q \times V_0$, $V_1' = Q \times V_1$, $v_0' = (q_0, v_0)$

- *For all $(q_1, v_1) \in V_0'$ and $a \in A_0$, we have $E_0'((q_1, v_1), a) = (q_1, E_0(v_1, a))$ if $L(q_1) = a$ and $E_0'((q_1, v_1), a)$ is undefined otherwise.*

- *For all $(q_1, v_1) \in V_1'$ and $a \in A_1$, we have $E_1'((q_1, v_1), a) = (\delta(q_1, a), E_1(v_1, a))$.*

- *For $\mathcal{F} = \{(B_1, C_1), \ldots, (B_n, C_n)\}$, $\mathcal{F}' = \{(Q \times B_1, Q \times C_1), \ldots, (Q \times B_n, Q \times C_n)\}$.*

*Proof.* The game built as described above essentially follows the paths induced by the strategy in the game graph without restricting the possible moves of player 1. Plays that are winning for player 0 $\mathcal{G}'$ correspond bijectively to winning plays for player 0 in $\mathcal{G}$ using the strategy $M$. □

We obtain as a summary:

**Corollary 9.** *Given some value $b$ and some game $\mathcal{G}$ with safety, Büchi, co-Büchi, generalised Büchi, parity, Rabin, Streett or Muller winning condition, the problem of deciding whether there exists a stand-alone or arena-based strategy of size $b$ in $\mathcal{G}$ that is winning for player 0 is contained in the complexity class NP (for a unary encoding of $b$ in the case of Streett or Muller two-player games).*

## 5 Arena-based strategies in games

There exists a theorem by Chatterjee et al. [4] that proves that there exists no polynomial-time approximation scheme (PTAS) for finding small positional strategies in safety games. As safety games are a special case of parity, Büchi, co-Büchi, generalised Büchi, Rabin and Streett games, the NP-hardness result also applies to these cases. In this paper, we strengthen the result that the problem cannot be approximated within **every** constant ($> 1$) in polynomial time to the fact that the problem cannot even be approximated within **any** constant in polynomial time. We follow the same reduction idea as in [4], but deviate by performing the hardness proving task by reducing the E$k$-Vertex-Cover problem onto finding small positional strategies.

**Definition 10** (E$k$-Vertex-Cover). *A $k$-uniform hypergraph is a 2-tuple $G = \langle V, E \rangle$ such that $V$ is a finite set and $E \subseteq 2^V$ such that all elements in $E$ are of cardinality $k$. Given a $k$-uniform hypergraph $H = \langle V, E \rangle$, the E$k$-Vertex-Cover problem is to find a subset $V' \subseteq V$ of minimal cardinality such that for all $e \in E$: $e \cap V' \neq \emptyset$.*

It has been proven that approximating the minimal size of an E$k$-Vertex-Cover within a factor of $(k-1-\epsilon)$ for some $\epsilon > 0$ is NP-hard [6].

**Theorem 11.** *For every $b \in \mathbb{N}$, the problem of approximating the size of a smallest arena-based winning strategy in a safety or Muller game within a factor of $b$ is NP-hard.*

*Proof.* We reduce the E$k$-Vertex-Cover problem $H = \langle V, E \rangle$ for $k = b+4$ to our case and construct a suitable game $\mathcal{G} = (V_0, V_1, A_0, A_1, E_0, E_1, v_0, \mathcal{F})$ as follows:

- $A_0 = V \cup \{\cdot\}$

- $A_1 = E \cup \{\cdot\}$
- $V_0 = \{v_0, \bot_0\} \cup E \cup V \times \{0, \ldots, |E|\}$
- $V_1 = \{v_1, \bot_1\} \cup V \times \{0, \ldots, |E|\}$
- $E_0(v_0, \cdot) = v_1$, $E_0(e, v) = (v, 0)$ for all $e \in E$ with $v \in e$, $E_0((v, j), \cdot) = (v, j+1)$ for all $v \in V$ and $j < |E|$, $E((v, |E|), \cdot) = \bot_1$ for all $v \in V$, $E_0(\bot_0, \cdot) = \bot_1$; for the remaining combinations of $v \in V_0$ and $a \in A_0$, $E_0(v, a)$ is undefined
- $E_1(v_1, e) = e$ for all $e \in E$, $E_1((v, j), \cdot) = (v, j)$ for all $v \in V$ and $0 \leq j \leq |E|$, $E_1(\bot_1, \cdot) = \bot_0$; for the remaining combinations of $v \in V_1$ and $a \in A_1$, $E_1(v, a)$ is undefined
- For the reduction to the Muller case, $\mathcal{F} = \{\bot_0\}$

Figure 1 gives an example of such a game. Intuitively, player 1 first chooses an edge and player 0 then selects a vertex that is an element of the edge. Since player 1 can choose any edge and a small arena-based strategy is to be found, player 0 has to choose her moves strategically wise in order to obtain a small strategy. The long slides for every vertex amplify the effects of bad decisions of player 0. Note that in both the safety and Muller game case, we can restrict our attention to positional strategies as the only vertices of the game that can be visited more than once along a path are $\bot_0$ and $\bot_1$, which form an absorbing strongly connected component.

There exists a one-to-one-correspondence for player 0 between winning strategies and vertex covers in $H = \langle V, E \rangle$. For every vertex cover with $j$ vertices, there exists a corresponding strategy with precisely $(1 + |E|) \cdot j + |E| + 2$ reachable $V_0$-positions. Now assume that there exists some factor-$b$ polynomial approximation algorithm for finding the smallest winning strategy. Then, for $V'$ being some E$k$-Vertex-Cover induced by the strategy found by the algorithm and $V'_{\text{opt}}$ being a smallest E$k$-Vertex-Cover in $H$ (without loss of generality, we can assume that the size of the minimal strategy is at least $b$ as such small solutions can be enumerated in polynomial time), we can deduce:

$$\frac{2 + |E| + (|E| + 1) \cdot |V'|}{2 + |E| + (|E| + 1) \cdot |V'_{\text{opt}}|} = \frac{1 + (|E| + 1) \cdot (|V'| + 1)}{1 + (|E| + 1) \cdot (|V'_{\text{opt}}| + 1)} \leq b \quad \Rightarrow \quad \frac{|V'|}{|V'_{\text{opt}}|} \leq b + 2 = k - 2$$

So the existence of a factor $b$-approximation algorithm for finding small positional strategies in a safety game would imply the existence of a polynomial $k - 2$-approximation algorithm for the E$k$-Vertex-Cover problem, which is a contradiction (unless P=NP). Since we can follow this line of reasoning for all $b \in \mathbb{N}$, the claim follows. □

In the set of winning conditions considered in this paper, the Rabin condition is a special case as games with this condition are NP-hard to solve even though they are memoryless determined. This fact can be exploited in order to obtain an even stronger non-approximability result for games with this winning condition.

**Corollary 12.** *For any strictly increasing function $g : \mathbb{N}_0 \to \mathbb{N}_0$, approximating the size of a smallest winning arena-based strategy in Rabin games within $g$ is NP-hard.*

*Proof.* This corollary is a direct consequence of the fact that we can reduce the problem of determining the winner in a $n$-state Rabin game to approximating the size of a smallest winning arena-based strategy within $g(n)$. □

## 6 Approximation algorithms for hard problems

In this section, we present a simple scheme for polynomial time algorithms that approximate the minimal winning strategy size in some game $\mathcal{G} = (V_0, V_1, A_0, A_1, E_0, E_1, v_0, \mathcal{F})$ within any exponential function $f(n) = c^n$ (with some $c > 1$) for some of the combinations of winning condition types and strategy types, provided that both $|A_0|$ and $|A_1|$ are fixed (and thus not taken into account for the complexity considerations).

In particular, the scheme is suitable for all cases in which (1) constructing some (not necessarily minimal) winning strategy is possible in polynomial time (whenever the game is winning for player 0) and (2) strategies that are exponentially smaller than the worst-case strategy size induced by the game type can be represented using an amount of memory that is logarithmic in the size of the game.

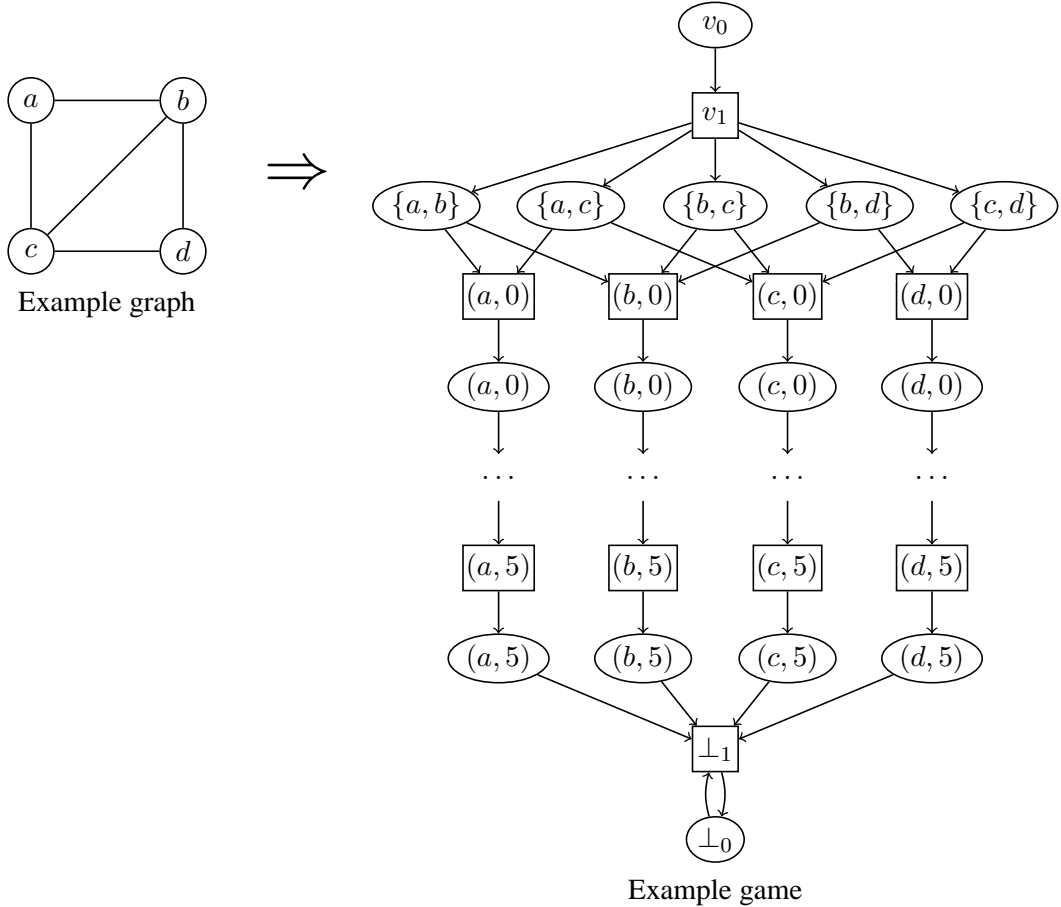

Figure 1: Example game construction from an instance of an E$k$-Vertex-Cover problem as described in the proof of Theorem 11. For simplicity, we have chosen $k = 2$ for this example. Positions of player $0$ are denoted by ellipses whereas rectangles mark the positions of player $1$. The action labels have been omitted.

Condition (1) holds in the case of safety, co-Büchi, Büchi, generalised Büchi and parity two-player games with a constant number of colours and also for all winning condition types dealt with here for one-player games.

Condition (2) holds for positional strategies in two-player games as well as for witnesses in one-player games.

The basic idea of the scheme is to first produce some initial strategy of size polynomial in the size of the game and then iterate over all exponentially smaller strategies and check if they are winning for player $0$.

As an example for the first step, for finding some positional strategy in a parity game with a constant number of colours, we can gradually remove positions and edges for player $0$ in the game in a random order and check after each removal whether the remaining game is still winning for player $0$ (which only needs polynomial time for a constant number of colours with current algorithms, see, e.g., [21]). Whenever this is not the case, we backtrack to the respective previous decision and continue with the next position or edge in the chosen order. When this process finishes, there is necessarily one winning deterministic strategy left. The currently fastest known automaton emptiness checking procedures for Streett automata (and thus, also for generalised Büchi automata as a special case) [19, 7] also produce such a certificate.

For finding small witnesses, we can first compute an accepting lasso and convert it to a witness by reading the actions along the lasso.

Then, after finding some initial strategy of size $n$, the algorithm proceeds by iterating over all strategies of size not more than $\lceil \log_c n \rceil$ and check for each of these whether they are winning. If the number of strategies of this size is only polynomial in the size of the original game structure and checking if a strategy is winning can be done in polynomial time (which is true for all types considered here), this steps takes only polynomial time (as the product of two polynomials is a polynomial). As whenever there exists a strategy of size not larger than $\lceil \log_c n \rceil$, the second step will find it and whenever this is not the case, we know that the smallest strategy must be of size between $n$ and $\lceil \log_c n \rceil$, we obtain the exponential approximation quality of this scheme.

It remains to show that in the cases considered above, Condition (2) actually holds, i.e., that the number of strategies of size $\lceil \log_c n \rceil$ is polynomial in $|\mathcal{G}|$.

In positional strategies, only one successor vertex has to be chosen for every reachable vertex of $V_0$, of which only $|A_0|$ many exist. Without loss of generality, we assume that the strategy is given in some arranged form (e.g., breadth-first order) such that we do not have to guess which states are reachable in the game. As a consequence, the total number of strategies is $(n+1) \cdot |A_0|^{\lceil \log_c n \rceil}$, which is clearly polynomial in $|\mathcal{G}|$.

For witnesses and non-positional lassos, we have to guess $\lceil \log_c n \rceil$ many elements of $|A_0|$ along with a splitting position between the first and the second part of the word. Thus, the total number of strategies is $\lceil \log_c n \rceil \cdot |A_0|^{\lceil \log_c n \rceil}$, which is clearly polynomial in $|\mathcal{G}|$.

As a result, we obtain:

**Corollary 13.** *Let $c > 1$ and $\Sigma$ be some fixed finite alphabet. Given some $\omega$-automaton $\mathcal{A} = (Q, \Sigma, q_0, \delta, \mathcal{F})$ with any of the acceptance types considered in this paper, computing a word $uv^\omega$ such that $|u| + |v|$ is not longer than $c^n$ for $n$ being the minimal witness length can be done in time polynomial in $|\mathcal{A}|$.*

**Corollary 14.** *Approximating the size of a smallest winning arena-based strategy for player $0$ in parity games with a constant number of colours (and their special cases) within any exponential function can be performed in time polynomial in the size of the game if the sets of actions for both players are fixed.*

# 7 Conclusion

In this paper, we have discussed the approximation hardness of the problem of obtaining small winning strategies, accepting lassos and witnesses in games and automata with $\omega$-regular winning or acceptance conditions. Some of the results obtained (including the ones from [9]) are surprising. For example, while finding short accepting lassos is doable in polynomial time even for the relatively complex Rabin condition, approximating the size of a shortest witness within any polynomial is NP-hard even for safety automata and thus considerably harder.

In some cases, the results obtained are tight. For example, we have characterised precisely the approximability of finding shortest witnesses in $\omega$-automata (with a constant-sized alphabet): the problem is NP-hard to approximate within every polynomial, but approximable in polynomial time within any exponential function. In other cases, the findings are only the first step in determining the precise approximation hardness of the problems, as for example in the two-player case for the Rabin, Streett and Muller winning condition types. Nevertheless, to the best of our knowledge, we presented the first comprehensive overview of the known results on the problem of obtaining small winning strategies, accepting lassos and witnesses in games and automata with $\omega$-regular winning or acceptance conditions, which will hopefully facilitate further research in this area.

From a theoretical point of view, our results show that size bounds imposed on certificates can significantly increase the complexity of tasks in verification and synthesis that employ computing such certificates. As a side remark, in these applications, the game is often represented in a symbolical form, like for example using reduced ordered binary decision diagrams [3, 26]. For such representations, the complexity of the problems discussed here might even be higher, but the analysis of such cases is beyond the scope of this paper.

From an algorithmic point of view, as the problems discussed here are of importance for many applications such as LTL and ATL model checking or synthesis of finite state systems, the non-approximability results level the ground for choosing the right approaches for solving these problems more effectively in the future. As for most cases, the approximability of the problems considered is quite bad, the practitioner can now conclude that looking into suitable heuristics might be worthwhile doing. For example, by encoding the problem into a SAT instance for applying modern SAT solvers, the problem could be tackled precisely. Automatic optimization of such solvers to this setting could further increase the performance (see, e.g., [15]). On the other hand, as for most applications, it suffices to obtain relatively small, but not necessarily smallest strategies, efficient heuristics might be more suitable for these tasks. Especially when it comes to symbolically represented game graphs, heuristic approaches that do not require representing the game graph in an explicit way appear to be promising.